# Fully 3D Monte Carlo image reconstruction in SPECT using functional regions

Ziad El Bitar[a,*], Delphine Lazaro[b], Christopher Coello[b]; Vincent Breton[a], David Hill[c] and Irène Buvat[b]

[a] *Laboratoire de Physique Corpusculaire, CNRS/IN2P3, Université Blaise Pascal, 24 Avenue des Landais, 63177 Aubière Cedex 1, France*
[b] *UMR 678 INSERM UPMC, CHU Pitié-Salpêtrière, 91 Boulevard de l'Hôpital, 75634 Paris Cedex 13, France*
[c] *ISIMA/LIMOS UMR 6158, Computer Science and Modeling Laboratory, Université Blaise Pascal, 24 Avenue des Landais, BP – 10125 ,63173 Aubière Cedex, France*

**Abstract**

Image reconstruction in Single Photon Emission Computed Tomography (SPECT) is affected by physical effects such as photon attenuation, Compton scatter and detector response. These effects can be compensated for by modeling the corresponding spread of photons in 3D within the system matrix used for tomographic reconstruction. The fully 3D Monte Carlo (F3DMC) reconstruction technique consists in calculating this system matrix using Monte Carlo simulations. The inverse problem of tomographic reconstruction is then solved using conventional iterative algorithms such as maximum likelihood expectation maximization (MLEM). Although F3DMC has already shown promising results, its use is currently limited by two major issues: huge size of the fully 3D system matrix and long computation time required for calculating a robust and accurate system matrix. To address these two issues, we propose to calculate the F3DMC system matrix using a spatial sampling matching the functional regions to be reconstructed. In this approach, different regions of interest can be reconstructed with different spatial sampling. For instance, a single value is reconstructed for a functional region assumed to contain uniform activity. To assess the value of this approach, Monte Carlo simulations have been performed using GATE. Results suggest that F3DMC reconstruction using functional regions improves quantitative accuracy compared to the F3DMC reconstruction method proposed so far. In addition, it considerably reduces disk space requirement and duration of the simulations needed to estimate the system matrix. The concept of functional regions might therefore make F3DMC reconstruction practically feasible.

*Keywords* : Single Photon Emission Computed Tomography; Monte Carlo simulation; Image reconstruction

## 1. Introduction

Single Photon Emission Computed Tomography (SPECT) is affected by several physical phenomena, such as photon attenuation, Compton scatter, and limited detector response function, which bias image accuracy if not compensated for. As all these phenomena are 3D in nature, the most relevant approach to account for them is to model their impact in 3D during the tomographic reconstruction process (e.g. [1, 2]). Although attenuation and detector response can be accurately modelled using analytical expressions (e.g. [3]), modelling Compton scatter is more of a challenge. Monte Carlo modelling has been proposed as a means of accurately describing all 3D phenomena involved in the SPECT imaging process [4, 5]. In this approach, the system matrix describing the forward model, i.e. the link between the actual activity distribution and the events collected by the detector, is calculated using Monte Carlo simulations, given the geometry and attenuation properties of the object under investigation. Tomographic reconstruction is then performed by inverting this system matrix using maximum likelihood estimation maximisation (MLEM). Although promising, the practical application of this fully 3D Monte Carlo (F3DMC) reconstruction is limited by two issues: duration of the simulations needed to calculate the system matrix, and storage and manipulation of the huge system matrix. As an example, a SPECT acquisition including 64 projections of 64 x 64 pixels yields a system matrix of about 69 gigabytes (using one byte to store a matrix element) if the object to be reconstructed is sampled on a 64 x 64 x 64 voxel grid.

To address these problems, we suggest to reconstruct the object using a spatial sampling that matches the functional regions present in the object, yielding less values to be reconstructed than when performing a voxel-by-voxel reconstruction. The system matrix can then be calculated using the same sampling, which considerably reduces its size, and also makes each element of the matrix less affected by noise than when calculated on a voxel-by-voxel basis. In section 2, the F3DMC reconstruction approach involving functional regions is described. Section 3 presents the methods used to test the concept, while results are shown in section 4. Finally, the method is discussed in section 5.

## 2. Theory

In SPECT, a discrete formulation of the problem of image reconstruction reads:

$$p = \mathbf{R} f \qquad (1)$$

where $p$ is a column vector of $M \times N \times P$ elements corresponding to $M$ projections each containing $N \times P$ projection bins, $f$ is a column vector of $X \times Y \times Z$ elements corresponding to the activity distribution to be reconstructed and $\mathbf{R}$ is the $(M \times N \times P, X \times Y \times Z)$ system matrix. Because of scatter and attenuation, elements of the system matrix $\mathbf{R}$ depend on the geometry and attenuation properties of the object to be reconstructed. This information can be obtained by a computed tomography (CT) scan. In F3DMC reconstruction, $\mathbf{R}$ is computed using Monte Carlo simulations, given the geometry and attenuation properties of the object, and assuming a uniform activity distribution within the object [4, 5]. Each element $\mathbf{R}_{ij}$ of $\mathbf{R}$ is defined as the probability that a photon emitted from voxel $j$ is detected within projection bin $i$ and is computed as:

$$\mathbf{R}_{ij} = \frac{N_{ij}}{N_j} \qquad (2)$$

where $N_{ij}$ is the number of photons emitted from voxel $j$ and detected in the projection bin $i$, and $N_j$ is the number of photons emitted from voxel $j$.

Once the system matrix R is computed, the MLEM algorithm is used to solve equation (1), because of the Poisson nature of SPECT projections [6].

To reduce the size of the system matrix, we propose not to describe the activity distribution $f$ to be reconstructed using a voxel-by-voxel sampling, but to represent it as a set of functional regions, each functional region being assumed to contain a uniform activity distribution. With this representation, $f$ can be fully described using $F$ values, corresponding to the activity concentration in each of the functional regions. $f$ is thus a column vector of $F$ elements instead of $X \times Y \times Z$ elements, and $\mathbf{R}$ becomes a $(M \times N \times P, F)$ matrix.

## 3. Materials and Methods

### 3.1. Phantom

A water cylinder (10 cm high, 10 cm in diameter) was considered, in which 6 rods of different diameters (4.8 mm, 6.4 mm, 7.8 mm, 9.6 mm, 11.1 mm and 12.7 mm) were included (Fig. 1). The biggest road contained bony material whereas the five smaller rods were filled with water. The cylinder and the rods were filled by 2.08 MBq/ml and 8.32 MBq/ml of Tc-99m respectively (rod-to-background activity concentration ratio of 4), while the bony rod contained no activity. The volume of the phantom to be reconstructed was sampled on a 64 x 64 x 64 voxel grid (voxel size = 3.125 mm x 3.125 mm x 3.125 mm).

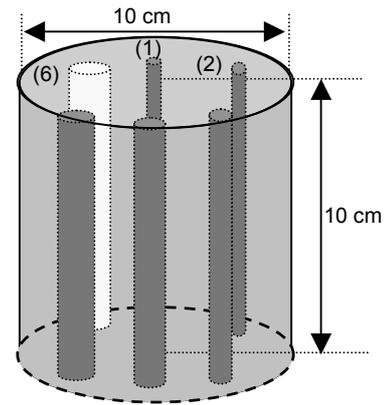

Fig. 1. Water cylinder phantom.

### 3.2. Simulation of a SPECT acquisition

Monte Carlo simulations were performed with GATE [7]. An AXIS Philips gamma camera equipped with an LEHR (Low Energy High Resolution) collimator was modeled. Data corresponding to a SPECT acquisition of 64 projections 64 x 64 (pixel size = 3.125 mm x 3.125 mm) were simulated. 29 billion photons were generated, among which 6.2 millions were detected within the 126-154 keV energy window.

### 3.3 Calculation of F3DMC system matrices

To calculate the system matrix, a uniform activity concentration was set in the phantom and a SPECT acquisition with identical parameters as described in section 3.2. was simulated. 74 billion events were simulated and 16 million were detected within the 126-154 keV energy window. Simulation duration was about 10 days on a cluster of 20 Pentium III 1GHz PC. From these simulated data, four types of system matrix were derived: the conventional F3DMC system matrix, a matrix involving functional regions, a hybrid matrix involving both voxels and functional regions, and matrices involving merged functional regions.

*Conventional F3DMC matrix.* The (262144 x 262144) system matrix **R** corresponding to a voxel-by-voxel representation of *f* was calculated. As only non-zero elements were stored, this matrix was actually about 88 Mbytes big.

*F3DMC matrix with functional regions.* Given the considered phantom, 7 functional regions were used, corresponding to the 6 rods and the complementary background region. The system matrix **R**$_F$ was therefore only (262144 x 7), and the reconstruction problem consisted in estimating 7 unknown values only, namely the activity concentration within each functional region. Each **R**$_{Fij}$ element of matrix **R**$_F$ was the probability that a photon emitted from functional region *j* is detected in projection bin *i*.

*Hybrid F3DMC matrix.* The hybrid approach assumes that part of the activity distribution can be segmented into functional regions with uniform activity, while other parts need to be represented using a voxel-by-voxel description. For our phantom, we considered the background as a functional region with uniform activity (one single value to be estimated for the background) while all other regions were reconstructed using a voxel-by-voxel sampling. The resulting system matrix **R**$_H$ included 262144 rows and 2497 columns. Each **R**$_{Hij}$ element of matrix **R**$_H$ was the probability that a photon emitted from functional region *j* (or from voxel j) is detected in projection bin *i*.

*F3DMC matrices involving merged functional regions.* System matrices **R**$_M$ involving "merged" functional regions were also calculated. To do that, the phantom was segmented into 3 regions only: one over one rod (called "rod of interest"), one in the center of the phantom including background only, and one including all other voxels (i.e., voxels belonging to the background and all other rods except the rod of interest). This was repeated 6 times, each with a different rod as the rod of interest.

Given these different system matrices, F3DMC reconstruction was completed using MLEM. The criteria used to stop iterations was that the sum of the squared differences between the estimated projection bins and the measured projection bin defined as:

$$Diff(n) = \sum_i \sum_j (\mathbf{R}_{ij} \times f_j^n - p_i)^2 \quad (3)$$

was smaller than 0.01. In equation (3), $f_j^n$ is the reconstructed image value in voxel j (or functional region j) and $p_i$ is the number of photons actually detected in projection bin *i*.

### 3.4. Figures of merit for evaluation

To compare the approaches discussed above, we calculated the rod-to-background activity concentration ratio for each rod and compared it with reference values. Although the theoretical values of these ratios should be 4 for all hot rods and 0 for the bony rod, the very sampling of the activity distribution used for Monte Carlo simulation introduced biases. Therefore, reference values were taken as the rod-to-background activity ratios that were measured on the image of emitted photons (Figure 2). These values were 1.92, 2.23, 2.46, 2.54, 2.73 and 0.41 going from the smallest to the biggest rod.

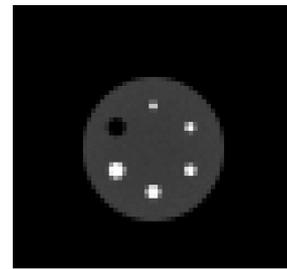

Fig. 2. Slice through the reference activity distribution.

### 4. Results

Table 1 shows the rod-to-background activity concentration ratios as measured in the images reconstructed with F3DMC using **R**, **R**$_F$, **R**$_H$ and **R**$_M$, together with the associated percent errors between estimated and true activity ratios. Note that the results using **R**$_M$ actually correspond to 6 reconstructions, each with a different system matrix (see section 3.3). This is why there are 6 numbers of iterations and 6 matrix sizes associated with the results.

It can be seen from this table that all **R**$_F$, **R**$_H$ and **R**$_M$ gave accurate quantitative results, which were less biased that those with **R**.

| | Rod 1 | Rod 2 | Rod 3 | Rod 4 | Rod 5 | Rod 6 | Number of iterations | Disk space (MBytes) |
|---|---|---|---|---|---|---|---|---|
| **R** | 1.65 (-16) | 2.01 (-11) | 2.15 (-14) | 2.42 (-5) | 2.66 (-2) | 0.52 (22) | 500 | 88.3 |
| **R$_F$** | 1.96 (2) | 2.34 (5) | 2.47 (0.6) | 2.69 (6) | 2.87 (5) | 0.43 (7) | 300 | 2.3 |
| **R$_H$** | 1.94 (1) | 2.34 (5) | 2.46 (0.2) | 2.52 (-0.5) | 2.95 (8.3) | 0.47 (5) | 2000 | 9.4 |
| **R$_M$** | 1.90 (-1) | 2.25 (1) | 2.46 (0.1) | 2.54 (0.04) | 2.69 (-2.37) | 0.4 (-1) | 16, 14, 11, 11, 6, 36 | 2.1, 2.3, 2.7, 3, 3.3, 3.8 |

Table 1. Rod-to-background activity concentration ratios and percent errors (in parenthesis) with respect to the reference values as a function of the system matrix used for F3DMC reconstruction. Also shown are the number of MLEM iterations and the disk space required for storing the system matrix.

Figure 3 shows the percent errors in activity ratios with respect to the reference values as a function of the number of simulated photons used to calculate the system matrix, for images reconstructed with **R** and **R$_F$**. Results are shown for the smallest and the largest rods only but similar results were observed for all rods. This graph shows that to achieve the certain level of accuracy, F3DMC reconstruction with **R$_F$** requires between 4 and 5 less photons for system matrix calculation than reconstruction with **R**.

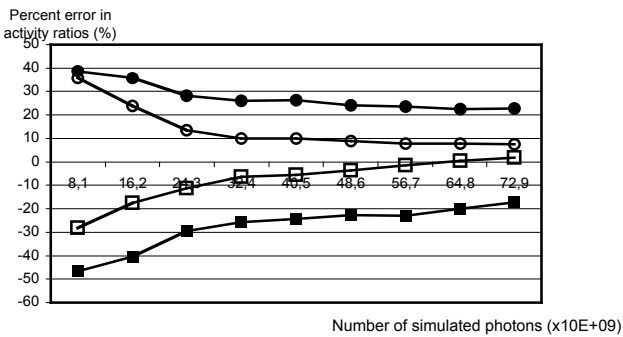

Fig 3. Percent error in activity concentration ratios with respect to the number of simulated photons. Circles : 12.7 mm rod, squares: 4.8 mm rod. Black symbols: **R** reconstruction, Open symbols: **R$_F$** reconstruction.

## 5. Discussion

### 5.1. Advantages of the use of functional regions

The major issues facing F3DMC reconstruction as initially proposed [4, 5] are the huge size of the system matrix and the large number of photons that need to be simulated to get a robust estimate of each probability value that the system matrix is composed of. The concept of functional regions brings solutions to these two problems. Indeed, the size of the system matrix is greatly reduced: instead of estimating activity values in each voxel, priors are used to identify voxels expected to have the same activity value (hence to belong to the same functional region), and a single activity value is estimated for all voxels belonging to the same functional region. The number of values to be estimated (number of components in $f$) is thus strongly reduced, yielding a concomitant reduction of the system matrix size, as the number of columns of this matrix equals the number of components in $f$. The disk space required to store the system matrix will obviously depend on the number of functional regions, and also on their size in the case of the hybrid approach, but will typically be much greater than 10. Reducing the system matrix size obviously also reduces the reconstruction time.

In addition, each element of the system matrix **R$_F$** is estimated using a larger number of photons than when using **R**. Indeed, each element **R$_{Fij}$** of the system matrix represents the probability $N_{ij}/N_j$ that a photon emitted from the functional region $j$ is detected in projection bin i (see equation 2). As each functional region $j$ includes a set of voxels, both $N_{ij}$ and $N_j$ are greater than when calculating **R** on a voxel-by-voxel basis, hence the uncertainty affecting **R$_{Fij}$** is smaller than that affecting **R$_{ij}$**. In our configuration, Figure 3 suggests that at least 4 times less photons are needed for system matrix calculation when using **R$_F$** instead of **R**, to achieve similar quantitative accuracy. This reduction of the number of photons to be simulated to calculate the system matrix will obviously depend on the number and size of functional regions.

### 5.2. Limits of the use of functional regions

Implementing this approach obviously requires strong priors to be able to describe the activity distribution as a set of functional regions, each characterized by a specific activity value. In real scans (unlike phantoms), it is very unlikely that the activity distribution can be simply represented by a set of easy-to-identify functional regions. This is why the simple idea of F3DMC reconstruction using functional regions was extended to the concept of F3DMC reconstruction using a hybrid system matrix. In this approach, the activity distribution to be reconstructed does not have to be segmented a priori into functional regions. Only part of it has to be identified as such. For instance, in a SPECT myocardial scan, lung activity could be assumed to be constant in each lung, and each lung could be considered as a functional region. On the other hand, one does not want to set any prior regarding the myocardial activity distribution which is of interest, hence the myocardial activity distribution could be reconstructed on a voxel-by-voxel basis. The hybrid approach is thus a trade-off between a voxel-by-voxel image reconstruction and a functional region reconstruction.

One way to define the functional regions needed for implementing F3DMC reconstruction using either functional regions or hybrid regions could be to perform a standard reconstruction first, using a conventional algorithm (filtered backprojection, or 2D OSEM for instance [8]), to help define regions in which activity could be set as uniform. Given these regions, reconstruction could be repeated using F3DMC functional or hybrid system matrix. The feasibility of the definition of relevant functional regions will be studied on real data in the future.

### 5.3. Comparative performance of the different F3DMC approaches

Table 1 shows that F3DMC reconstruction involving either $R_F$, $R_H$ and $R_M$ yielded more accurate images than when using $R$. Therefore, in addition to making F3DMC reconstruction practical by decreasing the size of the system matrix, the concept of functional regions improves quantitative accuracy. This is a direct consequence of the improved robustness of the matrix. As seen in Figure 3, similar accuracy could probably be achieved using $R$, but this would require at least 4 times more photons for calculating the system matrix than when using $R_F$.

Using $R_F$, $R_H$ and $R_M$, reconstructed images demonstrated a high degree of accuracy, with errors in activity concentration ratios less than 10%.

The definition of $R_M$ typically corresponds to a case where one is interested in a specific region in an image (the rod of interest in our phantom), and where other regions are not of particular interest. A "pure" background region was still considered (to allow for the calculation of a rod-to-background activity ratio), but all other voxels were merged into a single region. Results suggest that even when one merges several functional regions into a single region, activity can still be accurately estimated in another functional region well identified. This might have important applications. For instance, if specifically interested in a tumour uptake, one could define a functional region around the tumour, and merge all other voxels into few functional regions without having to perform a careful segmentation. Our phantom results suggest that in this case, the tumour activity would still be properly estimated. This will have to be confirmed using more complicated activity distributions.

### 5.4. Other applications

The concepts of functional regions and hybrid system matrix presented here in the context of SPECT could obviously be extended to PET. The principle of F3DMC reconstruction in PET is similar to that in SPECT (e.g., [9-10]), but the system matrix size is even more of an issue due to the large number of lines of response in PET. For that reason, only detector dependent effects are modelled in the system matrix (no modelling of object dependent effects such as scatter and attenuation), so that one can take advantage of some symmetry to reduce the system matrix size. Functional region and hybrid system matrix might therefore also be quite helpful for fully 3D reconstruction in PET.

### 6. Conclusion

F3DMC reconstruction using functional regions or hybrid system matrix offers a powerful solution to the issues of system matrix size and system matrix robustness faced by conventional F3DMC reconstruction. By reconstructing the object of interest using a sampling which matches the expected activity distribution, the system matrix size is greatly reduced and its robustness is improved with respect to the system matrix used in a voxel-by-voxel based F3DMC reconstruction. The resulting reconstructed images are hence more accurate, with errors in activity concentration ratio estimates less than 10%.

These new concepts make the application of F3DMC reconstruction on real data practical. The next step will be an experimental validation of the approach, first on phantom data acquired on a real camera, then on clinical data.